\documentclass[english,pra]{revtex4}
\usepackage[T1]{fontenc}
\usepackage[latin9]{inputenc}
\usepackage{amsmath}
\usepackage{graphicx}
\usepackage{amssymb}
\usepackage{color}
\usepackage{subfigure}
\usepackage{babel}
\usepackage{babel}

\begin{document}

\title{Spatial quadratic solitons guided by narrow layers of a nonlinear
material}
\author{Asia Shapira,$^{1,*}$ Noa Voloch-Bloch,$^1$ Boris A. Malomed,$^1$ and Ady Arie$^1$}
\address{$^1$Department of Physical Electronics, School of Electrical Engineering, Tel
Aviv University, Tel Aviv 69978, Israel}
\address{$^*$Corresponding author: asiasapi@post.tau.ac.il}

\begin{abstract}
We report analytical solutions for spatial solitons supported by layers
of a quadratically nonlinear ($\chi^{(2)}$) material embedded into
a linear planar waveguide. A full set of symmetric, asymmetric, and
antisymmetric modes pinned to a symmetric pair of the nonlinear layers
is obtained. The solutions describe a bifurcation of the \textit{subcritical}
type, which accounts for the transition from the symmetric to asymmetric
modes. The antisymmetric states (which do not undergo the bifurcation)
are completely stable (the stability of the solitons pinned to the
embedded layers is tested by means of numerical simulations). Exact
solutions are also found for nonlinear layers embedded into a nonlinear
waveguide, including the case when the uniform and localized $\chi^{(2)}$
nonlinearities have opposite signs (competing nonlinearities). For
the layers embedded into the nonlinear medium, stability properties
are explained by comparison to the respective cascading limit.
\end{abstract}

\maketitle

\section{Introduction}

The use of composite materials and engineered optical media opens
ways to new modes of the guided wave propagation, including self-trapped
nonlinear ones, in the form of spatial solitons. Recent reviews summarize
results obtained along these directions for photonic crystals \cite{PhotCryst}
and quasicrystals \cite{QuaCryst}, quasi-discrete media \cite{PhotLatt1},
and nonlinear lattices, which feature periodic modulation of the local
nonlinearity \cite{Barcelona}.

In ordinary settings, optical solitons are supported by uniform nonlinearities
(cubic, quadratic, or saturable), which may be combined with a periodic
grating, that plays the role of a linear potential of the lattice
type \cite{Wang}, and is necessary for stabilizing solitons against
the collapse in the multidimensional geometry \cite{J-Opt}. On the
other hand, spatially modulated nonlinearities may themselves induce
an effective potential \cite{Barcelona}. In particular, an interesting
issue is a possibility to support solitons by localized nonlinearities
embedded into linear host media. To introduce the topic, we will resort
here to a couple of simple models that admit analytical solutions,
and thus provide for a direct insight into specific properties of
solitons supported by the localized nonlinearity.

The simplest model of this type was introduced in Ref. \cite{Azbel},
in the form of the nonlinear Schredinger (NLS)\ equation for the
wave amplitude $u(x,z)$, with the cubic nonlinearity localized at
a single point: \begin{equation}
iu_{z}+(1/2)\psi_{xx}+\delta(x)|u|^{2}u=0.\label{Mark}\end{equation}
In terms of the optical transmission, $z$ is the propagation distance
and $x$ the transverse coordinate. Obviously, Eq. (\ref{Mark}) amounts
to a linear equation at $x<0$ and $x>0$, supplemented by the jump
condition for the derivative at $x=0$, which is produced by the integration
of the equation around $x=0$:\begin{equation}
~u_{x}\left(x=+0\right)-u_{x}\left(x=-0\right)=-2\left\vert u(x=0)\right\vert ^{2}u(x=0).\label{xx}\end{equation}
A family of exact solutions to Eq. (\ref{Mark}), in the form of \textit{peakons},
is obvious: \begin{equation}
u\left(x,z\right)=(2k)^{1/4}\exp\left(ikz-\sqrt{2k}|x|\right),\label{simple}\end{equation}
with arbitrary propagation constant $k>0$. This family features degeneracy,
as the power (norm) of the solutions does not depend on $k$, $P\equiv\int_{-\infty}^{+\infty}|u(x)|^{2}dx\equiv1$.
In particular, the formal application of the Vakhitov-Kolokolov (VK)
criterion, $dP/d\mu<0$, which is a necessary stability condition
for solitons in self-focusing nonlinear media \cite{VK}, predicts
neutral stability of solutions (\ref{simple}). In fact, all these
degenerate solitons are unstable, collapsing into a singularity or
decaying, as illustrated by another analytical solution to Eq. (\ref{Mark}),
which explicitly describes the onset of the collapse at $z\rightarrow-0$
\cite{Nir}: \begin{equation}
\psi\left(x,z\right)=\sqrt{-x_{0}/z}\exp\left[i\left(|x|-ix_{0}\right)^{2}/\left(2z\right)\right].\label{similariton}\end{equation}
Here $x_{0}>0$ is an arbitrary real constant, and $z$ is negative.
The same solution (\ref{similariton}) with $x_{0}<0$ describes decaying
solitons at $z>0$ \cite{Nir}. The power of this solution is also
$P=1$, irrespective of the value of $x_{0}$.

The solitons may be stabilized if a linear periodic potential is added
to Eq. (\ref{Mark}) \cite{Nir}. The stability is also achieved if
the single $\delta$-function in Eq. (\ref{Mark}) is replaced by
a symmetric pair, which corresponds to the equation introduced in
Ref. \cite{Dong}, \begin{equation}
iu_{z}+(1/2)\psi_{xx}+\left[\delta(x-L/2)+\delta(x+L/2)\right]|\psi|^{2}\psi=0.\label{2delta}\end{equation}
Exact analytical solutions to Eq. (\ref{2delta}) were found in Ref.
\cite{Dong} for symmetric, antisymmetric and asymmetric localized
modes. The respective, spontaneous-symmetry-breaking (SSB) bifurcation,
which generates asymmetric solutions from the symmetric ones, takes
place, with the increase of the power, at its critical value $P_{\mathrm{cr}}=\left(8/9\right)\left[1+\left(1/3\right)\ln2\right]\approx0.95$.
In this model, based on the ideal $\delta$-functions, the bifurcation
is degenerate, featuring an \textit{ultimately subcritical} character:
branches of the asymmetric solutions go backward as functions of $P$,
up to the state, attained at $P=1$, in which the entire power is
concentrated in an infinitely narrow soliton pinned to one of the
two $\delta$-functions. Accordingly, these branches are fully unstable.
The symmetric modes are stable at $P<P_{\mathrm{cr}}$ and unstable
at $P>P_{\mathrm{cr}}$, while antisymmetric modes are completely
unstable, although they do not undergo any bifurcation.

The degenerate character of the model with the two ideal $\delta$-functions
is the price paid for its analytical solvability. The degeneracy is
lifted if the $\delta$-functions in Eq. (\ref{2delta}) are approximated
by regular expressions:\begin{equation}
\delta\left(x\mp L/2\right)\rightarrow\delta\left(x\mp L/2\right)\equiv\frac{1}{\sqrt{\pi}a}\exp\left(-\frac{\left(x\mp L/2\right)^{2}}{a^{2}}\right),\label{a}\end{equation}
with small regularizing parameter $a$. The numerical analysis of
the regularized model demonstrates that the branches of asymmetric
states \emph{turn forward} at some $P$, which causes the stabilization
of the asymmetric solutions past the turning points. At $a>a_{0}\approx0.2$,
the SSB bifurcation becomes \textit{supercritical}, i.e., the branches
of the asymmetric solutions go forward immediately after they emerge,
being completely stable \cite{Dong}.

Rather than being represented by a single spot or a symmetric pair,
as outlined above, the localized nonlinearity may be extended to a
periodic lattice of $\delta$-functions embedded into the linear medium.
The description of stationary modes in such a model can be exactly
reduced to stationary solutions of the discrete NLS equation \cite{Canberra2,Kominis},
which has been studied in detail \cite{NLS}. However, the periodic
nonlinearity does not admit asymmetric modes.

A fundamental role in optics belongs to second-harmonic generating
systems based on the quadratic ($\chi^{(2)}$) nonlinearity \cite{Buryak,Jena}.
In this connection, it is relevant to consider media with one or several
narrow $\chi^{(2)}$ layers embedded into a linear planar waveguide.
For the single layer approximated by the respective $\delta$-function,
exact solutions in the form of peakons, similar to those given by
Eq. (\ref{simple}), were found in Ref. \cite{Canberra}. Unlike solutions
(\ref{simple}), they are not degenerate (the total power depends
on the propagation constant), a bigger part of the solution family
being stable.

A new problem, which is considered in the present work, is to find
double peakons pinned to a \textit{symmetric pair} of $\chi^{(2)}$
delta-functions, cf. Eq. (\ref{2delta}) for the $\chi^{(3)}$ nonlinearity.
In this model, we report analytical solutions of all the types, \textit{viz}.,
symmetric, antisymmetric (as concerns the fundamental-frequency component),
and asymmetric ones. The corresponding SSB bifurcation is subcritical,
but nongenerate (i.e., asymmetric branches eventually turn forward
as stable ones). Antisymmetric modes do not undergo bifurcations,
and turn out to be stable.

Another new configuration is a nonlinear \textit{double layer }(alias
a dipolar layer), formed by a fused pair of two narrow nonlinear stripes
with opposite signs. While it would be very difficult to create such
a configuration for the Kerr nonlinearity, in $\chi^{(2)}$ systems
it is more feasible, as the sign of the nonlinearity may be changed
by reversing the orientation of ferroelectric domains accounting for
the $\chi^{(2)}$ interaction. We consider the double layer described
by function $\delta^{\prime}(x)$ in front of the $\chi^{(2)}$ terms.
A family of exact solutions for solitons pinned to the double layer
is found, but they all turn out to be unstable (throughout the paper,
the stability is tested via direct simulations of the perturbed evolution,
in the framework of equations with the ideal $\delta$-functions replaced
by their regularized counterparts).

The most challenging problem is to construct analytical solutions
for a soliton pinned to a nonlinear layer embedded into a \emph{nonlinear}
waveguide. In this setting, the signs of the localized and uniform
$\chi^{(2)}$ nonlinearities may be identical or opposite. We produce
exact solutions of two different types for this case, and test their
stability. We also find some particular exact symmetric solutions
for a pair of nonlinear layers inserted into the nonlinear waveguide.

The paper is structured as follows. In Section 2, we briefly recapitulate
the peakon solution for the single $\chi^{(2)}$ layer embedded into
the linear medium. In particular, we apply the adiabatic approximation
to the description of peakons pinned to the layer whose strength slowly
varies along the propagation distance. The most essential results
are reported in Section 3, dealing with the pair of $\chi^{(2)}$
layers embedded into the linear medium, including exact solutions
for asymmetric double peakons. Results for the double layer are presented
in Section 4, and the nonlinear layer(s) buried into the nonlinear
waveguide are considered in Section 5. The paper is concluded by Section
6.

\section{The $\chi^{(2)}$ monolayer embedded into the linear medium}

The basic model, with a single narrow channel carrying the $\chi^{(2)}$
nonlinearity, can be written as follows:

\begin{eqnarray}
i\frac{\partial A_{1}}{\partial Z}+\frac{1}{2k_{1}}\frac{\partial^{2}A_{1}}{\partial X^{2}}+\kappa\delta(X/k_{1})A_{2}A_{1}^{\ast}e^{-i\triangle kZ} & = & 0,\label{A1}\\
i\frac{\partial A_{2}}{\partial Z}+\frac{1}{2k_{2}}\frac{\partial^{2}A_{2}}{\partial X^{2}}+\kappa\delta(X/k_{1})A_{1}^{2}e^{i\triangle kZ} & = & 0,\label{A2}\end{eqnarray}
where $A_{1}$ and $A_{2}$ are local amplitudes of the fundamental-frequency
(FF) and second-harmonic (SH) components, $k_{1}$ and $k_{2}$ are
the respective wave numbers, $\kappa$ is the nonlinearity coefficients,
and $\triangle k$ the phase mismatch. To reduce the number of control
parameters, we transform Eqs. (\ref{A1}) and (\ref{A2}): $A_{1}(X,Z)\equiv(1/2)u(x,z)$,
$A_{2}(X,Z)\equiv v(x,z)e^{i\triangle kZ}$, and rescale the spatial
coordinates and coefficients by means of $k_{1}$, $x\equiv X/k_{1}$,
$z\equiv Z/k_{1}$, $\gamma=\kappa/k_{1}$, and $Q=2\triangle k/k_{1}$.
The resulting normalized equations are

\begin{eqnarray}
iu_{z}+(1/2)u_{xx}+\gamma\delta(x)u^{\ast}v & = & 0,\label{u}\\
2iv_{z}+(1/2)v_{xx}-Qv+(\gamma/2)\delta(x)u^{2} & = & 0,\label{v}\end{eqnarray}
 The integration of the equations in an infinitesimal vicinity of
$x=0$ yields the relations for the jumps of gradients of the FF and
SH fields at $x=0$, cf. Eq. (\ref{xx}), \begin{eqnarray}
u_{x}(x & = & +0)-u_{x}(x=-0)=-2\gamma u^{\ast}(x=0)v(x=0),\label{ju}\\
v_{x}(x & = & +0)-v_{x}(x=-0)=-\gamma u^{2}(x=0),\label{jv}\end{eqnarray}
 while the fields themselves must be continuous across $x=0$.

A family of exact stationary solutions to Eqs. (\ref{u}) and (\ref{v})
in the form of peakons, similar to solution (\ref{simple}), is \cite{Canberra}\begin{eqnarray}
u_{\mathrm{peak}}\left(z,x\right) & = & \pm2\left[k\left(4k+Q\right)\right]^{1/4}\gamma^{-1}e^{ikz}\exp\left(-\sqrt{2k}|x|\right),\nonumber \\
v_{\mathrm{peak}}\left(z,x\right) & = & \sqrt{2k}\gamma^{-1}e^{2ikz}\exp\left(-\sqrt{2\left(4k+Q\right)}|x|\right).\label{exact}\end{eqnarray}
Equations (\ref{u}) and (\ref{v}) conserve the total power, alias
the \textit{Manley-Rowe invariant},\begin{equation}
P=\int_{-\infty}^{+\infty}\left[\left\vert u(x)\right\vert ^{2}dx+4\left\vert v(x)\right\vert ^{2}\right]dx\label{P}\end{equation}
For peakon (\ref{exact}), its value is\begin{equation}
P_{\mathrm{peakon}}=2\sqrt{2}\left(6k+Q\right)\left[\gamma^{2}\sqrt{4k+Q}\right]^{-1}.\label{T}\end{equation}
For $Q<0$, this dependence $P(k)$ has a positive slope, $dP/dk>0$,
at $k>-Q/3$, and a negative slope at $0<k<-Q/3$. According to the
VK\
criterion, the peakons should be stable for $k>-Q/3$, and unstable
for $0<k<-Q/3$. For $Q>0$, condition $dP/dk>0$ holds for all $k>0$,
hence the entire peakon family is expected to be \emph{stable} for
$Q>0$ \cite{Canberra}. It can be verified by direct simulations
that the stability of the peakons precisely complies with the predictions
of the VK criterion. Note also that expression (\ref{T}) gives rise
to a \textit{power threshold}: the solitons exist if their total power
exceeds a minimum value, which vanishes only at $Q=0$,\begin{equation}
P_{\min}=\left\{ \begin{array}{c}
P(k=0)\equiv2\sqrt{2Q}/\gamma^{2}~\mathrm{for}~~Q>0,\\
P(k=-Q/3)\equiv2\sqrt{-6Q}/\gamma^{2}~\mathrm{for}~~Q<0.\end{array}\right.\label{min}\end{equation}

In the case when the nonlinearity strength slowly varies along the
$\chi^{(2)}$ layer, i.e., $\gamma=\gamma(z)$ in Eqs. (\ref{u})
and (\ref{v}), the adiabatic approximation may be applied, assuming
that the solution given by Eqs. (\ref{exact}) remains locally valid
at each value of $z$, with slowly varying $k(z)$ which is determined
by the conservation of the total power, $P=\mathrm{const}$. Then,
one immediately finds from Eq. (\ref{T})\begin{equation}
k(z)=\frac{1}{4}\left\{ \left[\sqrt{\frac{P^{2}\gamma^{4}(z)}{72}+\frac{Q}{3}}+\frac{P\gamma^{2}(z)}{6\sqrt{2}}\right]^{2}-Q\right\} .\label{k}\end{equation}
There are limitations on the use of the adiabatic approximation: as
seen from Eq. (\ref{k}), for $Q<0$ this equation makes sense if
the expression under the radical is positive, i.e., $\gamma^{4}(z)>24|Q|/P^{2}.$
For $Q>0$, the radical is always real; however, in that case there
is another constraint, $k>0$ {[}solution (\ref{exact}) does not
make sense for $k<0${]}. As follows from Eq. (\ref{k}), this constraint
amounts to $\gamma^{4}(z)>8Q/P^{2}.$ Thus, the adiabatic approximation
does not allow the peakon to pass points where $\gamma(z)$ vanishes.

\section{The symmetric pair of nonlinear layers in the linear host medium}

\subsection{Formulation}

The modification of Eqs. (\ref{u}) and (\ref{v}) for two parallel
layers is obvious, cf. Eq. (\ref{2delta}):

\begin{eqnarray}
iu_{z}+\frac{1}{2}u_{xx}+\gamma\left[\delta\left(x-\frac{L}{2}\right)+\delta\left(x+\frac{L}{2}\right)\right]u^{\ast}v & = & 0,\label{2u}\\
2iv_{z}+\frac{1}{2}v_{xx}-Qv+\frac{\gamma}{2}\left[\delta\left(x-\frac{L}{2}\right)+\delta\left(x+\frac{L}{2}\right)\right]u^{2} & = & 0.\label{2v}\end{eqnarray}
Being a novel model for the $\chi^{(2)}$ nonlinearity, it is related
to its counterpart (\ref{2delta}) with the cubic nonlinearity through
the \textit{cascading limit}, which corresponds to large positive
values of mismatch $Q$ \cite{Jena,Buryak}. In this limit, one can
eliminate the SH field, using Eq. (\ref{2v}), $v\approx\left(\gamma/2Q\right)\left[\tilde{\delta}\left(x-L/2\right)+\tilde{\delta}\left(x+L/2\right)\right]u^{2},$
where it is necessary to assume that the ideal $\delta$-function
is replaced by its regularization $\tilde{\delta}(x)$, see Eq. (\ref{a}).
Then, the substitution of this approximation into Eq. (\ref{2u})
leads to equation\begin{equation}
iu_{z}+\frac{1}{2}u_{xx}+\frac{\gamma^{2}}{2Q}\left[\tilde{\delta}^{2}\left(x-\frac{L}{2}\right)+\tilde{\delta}^{2}\left(x+\frac{L}{2}\right)\right]|u|^{2}u=0.\label{limit}\end{equation}
With the ideal $\delta$-function, Eq. (\ref{limit}) does not make
sense, as $\delta^{2}(x)$ does not exist. Nevertheless, if $\tilde{\delta}\left(x\mp L/2\right)$
are taken as smooth approximations, Eq. (\ref{limit}) is meaningful,
being tantamount to the accordingly regularized version of Eq. (\ref{2delta}).

\subsection{General analysis}

Stationary localized solutions to Eqs. (\ref{2u}) and (\ref{2v})
are sought for as\begin{equation}
u(x,z)=e^{-ikz}\left\{ \begin{array}{c}
A_{-}e^{\sqrt{2k}\left(x+L/2\right)},~~\mathrm{at}~~x<-L/2,\\
A_{1}\cosh\left(\sqrt{2k}x\right)+A_{2}\sinh\left(\sqrt{2k}x\right),~~\mathrm{at}~~-L/2<x<+L/2,\\
A_{+}e^{-\sqrt{2k}\left(x-L/2\right)},~~\mathrm{at}~~x>L/2,\end{array}\right.\label{genu}\end{equation}
\begin{equation}
v(x,z)=e^{-2ikz}\left\{ \begin{array}{c}
B_{-}e^{\sqrt{2\left(4k+Q\right)}\left(x+L/2\right)},~~\mathrm{at}~~x<-L/2,\\
B_{1}\cosh\left(\sqrt{2\left(4k+Q\right)}x\right)+B_{2}\sinh\left(\sqrt{2\left(4k+Q\right)}x\right),~~\mathrm{at}~~-L/2<x<+L/2,\\
B_{+}e^{-\sqrt{2\left(4k+Q\right)}\left(x-L/2\right)},~~\mathrm{at}~~x>L/2.\end{array}\right.\label{genv}\end{equation}
Conditions of the continuity of $u$ and $v$ at $x=\pm L/2$ make
it possible to express the inner amplitudes, $A_{1,2}$ and $B_{1,2}$,
in terms of the outer ones, $A_{\pm}$ and $B_{\pm}$\begin{eqnarray}
A_{1} & = & \frac{A_{+}+A_{-}}{2\cosh\left(\sqrt{k/2}L\right)},~A_{2}=\frac{A_{+}-A_{-}}{2\sinh\left(\sqrt{k/2}L\right)},\nonumber \\
B_{1} & = & \frac{B_{+}+B_{-}}{2\cosh\left(\sqrt{\left(4k+Q\right)/2}L\right)},~B_{2}=\frac{B_{+}-B_{-}}{2\sinh\left(\sqrt{\left(4k+Q\right)/2}L\right)}~.\label{AABB}\end{eqnarray}
Further, we introduce notations\begin{eqnarray}
s_{1} & = & \left[\sinh\left(\sqrt{2k}L\right)\right]^{-1},~s_{2}=\left[\sinh\left(\sqrt{2\left(4k+Q\right)}L\right)\right]^{-1},\nonumber\\
c_{1} & = & 1+\coth\left(\sqrt{2k}L\right),~c_{2}=1+\coth\left(\sqrt{2\left(4k+Q\right)}L\right)~,\label{sc}\\
\gamma_{1} & = & \gamma/\sqrt{2k},~\gamma_{2}=\gamma/\sqrt{2\left(4k+Q\right)}~.\nonumber\end{eqnarray}
Then, conditions (\ref{ju}) and (\ref{jv}) for the jump of the wave
functions at points $x=\pm L/2$ give rise to the following equations:\begin{eqnarray}
c_{1}A_{-}-s_{1}A_{+} & = & 2\gamma_{1}A_{-}B_{-},~c_{1}A_{+}-s_{1}A_{-}=2\gamma_{1}A_{+}B_{+},\label{Ajump}\\
c_{2}B_{-}-s_{2}B_{+} & = & \gamma_{2}A_{-}^{2},~c_{2}B_{+}-s_{2}B_{-}=\gamma_{2}A_{+}^{2}~.\label{Bjump}\end{eqnarray}
Using Eqs. (\ref{Bjump}), one can eliminate $B_{\pm}$ in favor of
$A_{\pm}$:\begin{equation}
B_{+}=\frac{\gamma_{2}\left(c_{2}A_{+}^{2}+s_{2}A_{-}^{2}\right)}{c_{2}^{2}-s_{2}^{2}},~B_{-}=\frac{\gamma_{2}\left(c_{2}A_{-}^{2}+s_{2}A_{+}^{2}\right)}{c_{2}^{2}-s_{2}^{2}}~.\label{BB}\end{equation}

\subsection{Symmetric, asymmetric, and antisymmetric modes}

Substituting Eqs. (\ref{BB}) into Eqs. (\ref{Ajump}) and assuming
$A_{+}^{2}-A_{-}^{2}\neq0$ (i.e., that the solution is \emph{asymmetric})
leads to the following expressions for the amplitudes of the asymmetric
modes:\begin{equation}
A_{+}A_{-}=\frac{\left(c_{2}+s_{2}\right)s_{1}}{2\gamma_{1}\gamma_{2}},~A_{+}^{2}+A_{-}^{2}=\frac{c_{1}}{2c_{2}}\frac{c_{2}^{2}-s_{2}^{2}}{\gamma_{1}\gamma_{2}}~.\label{asymm}\end{equation}
The asymmetry of the stationary mode is characterized by \begin{equation}
\epsilon\equiv\frac{\left(A_{+}-A_{-}\right)^{2}}{\left(A_{+}+A_{-}\right)^{2}}=\frac{c_{1}c_{2}-c_{1}s_{2}-c_{2}s_{1}}{c_{1}c_{2}-c_{1}s_{2}+c_{2}s_{1}}.\label{epsilon}\end{equation}

On the other hand, Eqs. (\ref{Ajump}) and (\ref{Bjump}) immediately
give rise to the symmetric solutions,\begin{eqnarray}
A_{+} & = & A_{-}\equiv\left(A_{\pm}\right)_{\mathrm{symm}}=\sqrt{\frac{\left(c_{1}-s_{1}\right)\left(c_{2}-s_{2}\right)}{2\gamma_{1}\gamma_{2}}},\nonumber\\
B_{+} & = & B_{-}\equiv\left(B_{\pm}\right)_{\mathrm{symm}}=\left(2\gamma_{1}\right)^{-1}\left(c_{1}-s_{1}\right)~.\label{symm}\end{eqnarray}
Then, Eqs. (\ref{AABB}) yield, for the symmetric solution, $\left(A_{2}\right)_{\mathrm{symm}}=\left(B_{2}\right)_{\mathrm{symm}}=0,$
and\begin{equation}
\left(A_{1}\right)_{\mathrm{symm}}=\frac{\left(A_{\pm}\right)_{\mathrm{symm}}}{\cosh\left(\sqrt{k/2}L\right)},~\left(B_{1}\right)_{\mathrm{symm}}=\frac{\left(B_{\pm}\right)_{\mathrm{symm}}}{\cosh\left(\sqrt{\left(4k+Q\right)/2}L\right)}.\label{symm2}\end{equation}

The asymmetric solutions emerge, with the increase of $L$, from the
symmetric one as a result of the SSB bifurcation, at the point at
which the asymmetric solution, as given by Eqs. (\ref{asymm}), coincides
with its symmetric counterpart (\ref{symm}). Setting, accordingly,
$\epsilon=0$ in Eq. (\ref{epsilon}) predicts the location of the
bifurcation point: $c_{1}c_{2}=c_{1}s_{2}+c_{2}s_{1}.$ Only the symmetric
solution, given by Eqs. (\ref{symm}) and (\ref{symm2}), exists at
\begin{equation}
c_{1}c_{2}<c_{1}s_{2}+c_{2}s_{1}~,\label{no}\end{equation}
where Eq. (\ref{epsilon}) formally yields $\left(A_{+}-A_{-}\right)^{2}<0$,
while both the symmetric and asymmetric solutions exist at \begin{equation}
c_{1}c_{2}>c_{1}s_{2}+c_{2}s_{1}~,\label{yes}\end{equation}
when Eq. (\ref{epsilon}) yields $\left(A_{+}-A_{-}\right)^{2}>0$.

As follows from definitions (\ref{sc}), condition (\ref{no}) holds
in the limit when the two layers merge into one, $L\rightarrow0,$
in which case $c_{1}\approx s_{1}$ and $c_{2}\approx s_{2}$. On
the other hand, at $L\rightarrow\infty$ definitions (\ref{sc}) imply
$c_{1}\approx c_{2}\approx2$ and $s_{1,2}\rightarrow0,$ i.e., condition
(\ref{yes}) holds in the limit of a very large separation between
the layers. Therefore, the SSB indeed occurs, with the increase of
$L$, at a particular value of the separation.

It is also possible to find solutions which are antisymmetric in the
FF component, with $A_{+}=-A_{-}$ and $B_{+}=+B_{-}$, while the
SH component remains symmetric. In this case, Eqs. (\ref{AABB}),
(\ref{Ajump}), and (\ref{Bjump}) yield\begin{eqnarray}
A_{+} & = & -A_{-}\equiv\left(A_{\pm}\right)_{\mathrm{antisymm}}=\sqrt{\frac{\left(c_{1}+s_{1}\right)\left(c_{2}-s_{2}\right)}{2\gamma_{1}\gamma_{2}}},\nonumber\\
B_{+} & = & B_{-}\equiv\left(B_{\pm}\right)_{\mathrm{antisymm}}=\left(2\gamma_{1}\right)^{-1}\left(c_{1}+s_{1}\right)~.\label{anti}\end{eqnarray}
\begin{equation}
\left(A_{2}\right)_{\mathrm{antisymm}}=\frac{\left(A_{\pm}\right)_{\mathrm{antisymm}}}{\sinh\left(\sqrt{k/2}L\right)},~\left(B_{1}\right)_{\mathrm{antisymm}}=\frac{\left(B_{\pm}\right)_{\mathrm{antisymm}}}{\cosh\left(\sqrt{\left(4k+Q\right)/2}L\right)},\label{anti2}\end{equation}
and $\left(A_{1}\right)_{\mathrm{antisymm}}=\left(B_{2}\right)_{\mathrm{antisymm}}=0$,
cf. Eqs. (\ref{symm}) and (\ref{symm2}).

\subsection{Numerical results}

The numerical analysis of the dual-layer model aimed to address two
issues: the form of the bifurcation diagram (subcritical or supercritical),
which is implicitly described by the above analytical expressions,
and the stability of the symmetric, asymmetric, and antisymmetric
double peakons.

We took values of rescaled constants and variables which correspond
to $k=1$ and the following typical values of physical parameters:
at the FF wavelength of $1.0645~\mathrm{\mu}$m, the sample was assumed
to be stoichiometric lithium tantalate (SLT) with the e-ee interaction,
where two extraordinary waves induce an extraordinary nonlinear polarization.
Undoing the rescalings which lead to the rescaled notation, it is
straightforward to see that rescaled mismatch $Q=1$ corresponds,
in physical units, to a very small value, $6$ m$^{-1}$, i.e., our
actual results corresponds to the nearly matched $\chi^{(2)}$ system.

Further, the refractive indices and the relevant element of the $\chi^{(2)}$
susceptibility tensor were taken according to Ref. \cite{Dolev} (at
$100^{\circ}C$, $n_{\mathrm{FF}}=2.1323$, $n_{\mathrm{SH}}=2.1999$
and $d_{33}=12.9~$pm/V). In the simulations of the evolution of perturbed
solutions, the $\delta$-functions were replaced by approximation
(\ref{a}) with $a\sim$ a few microns. Finally, taking into regard
the experimentally measured value of the Kerr coefficient in this
material, $n_{2}\approx14.6\times10^{-16}$ cm$^{2}$/W \cite{Japan},
it is easy to check that, for the physical parameters adopted in this
work, the $\chi^{(3)}$ nonlinearity is negligible in comparison with
the $\chi^{(2)}$ effects.

\begin{figure}[tbp]
\includegraphics[width=0.5\textwidth]{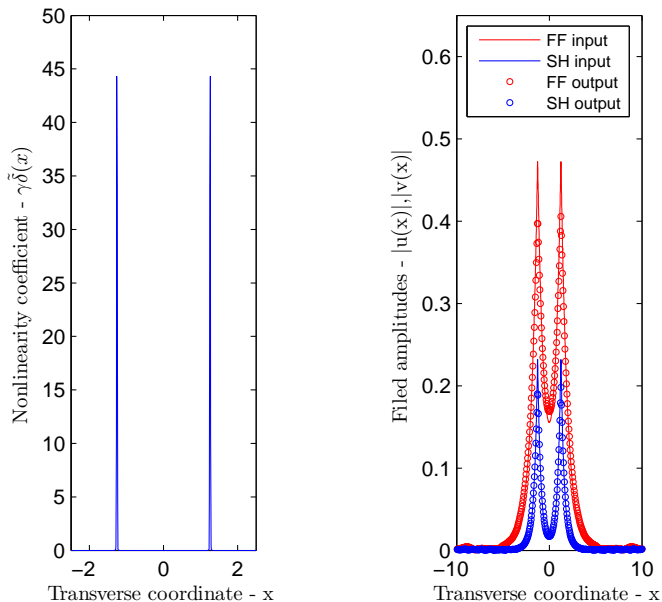}
\caption{(Color online) Left: the double-barrier structure corresponding, in
physical units, to separation $L=200$ $\mathrm{\protect\mu }$m between the
two symmetric $\protect\chi ^{(2)}$ layers of width $a=0.92~\mathrm{\protect%
\mu }$m each. Right: a typical example of stable symmetric solitons. The red
(taller) and blue (lower) continuous curves depict, respectively, the input
for the FF and SH fields, taken as per analytical solution given by Eqs. (%
\protect\ref{genu})-(\protect\ref{BB}) and (\protect\ref{symm}). Chains of
dots depict the output produced by simulations of Eqs. (\protect\ref{2u}), (%
\protect\ref{2v}), with the $\protect\delta $-functions approximated as per
Eq. (\protect\ref{a}), over the propagation distance corresponding to $z=100$
cm. For this and other examples, the scaled wavenumber of the analytical
solutions is taken as $k=1$. In the present case, the mismatch is $Q=0$.}
\label{fig1}
\end{figure}

\begin{figure}[tbp]
\includegraphics[width=0.5\textwidth]{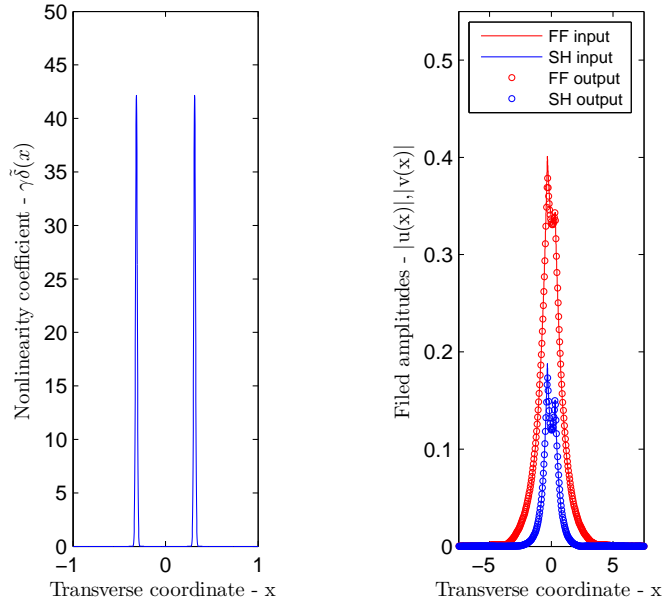}
\caption{(Color online) The same as in Fig. \protect\ref{fig1}, but for a
stable asymmetric soliton, with $L=50$ $\mathrm{\protect\mu }$m, $a=1.06~%
\mathrm{\protect\mu }$m, and $z=10$ cm.}
\label{fig2}
\end{figure}

\begin{figure}[tbp]
\includegraphics[width=0.5\textwidth]{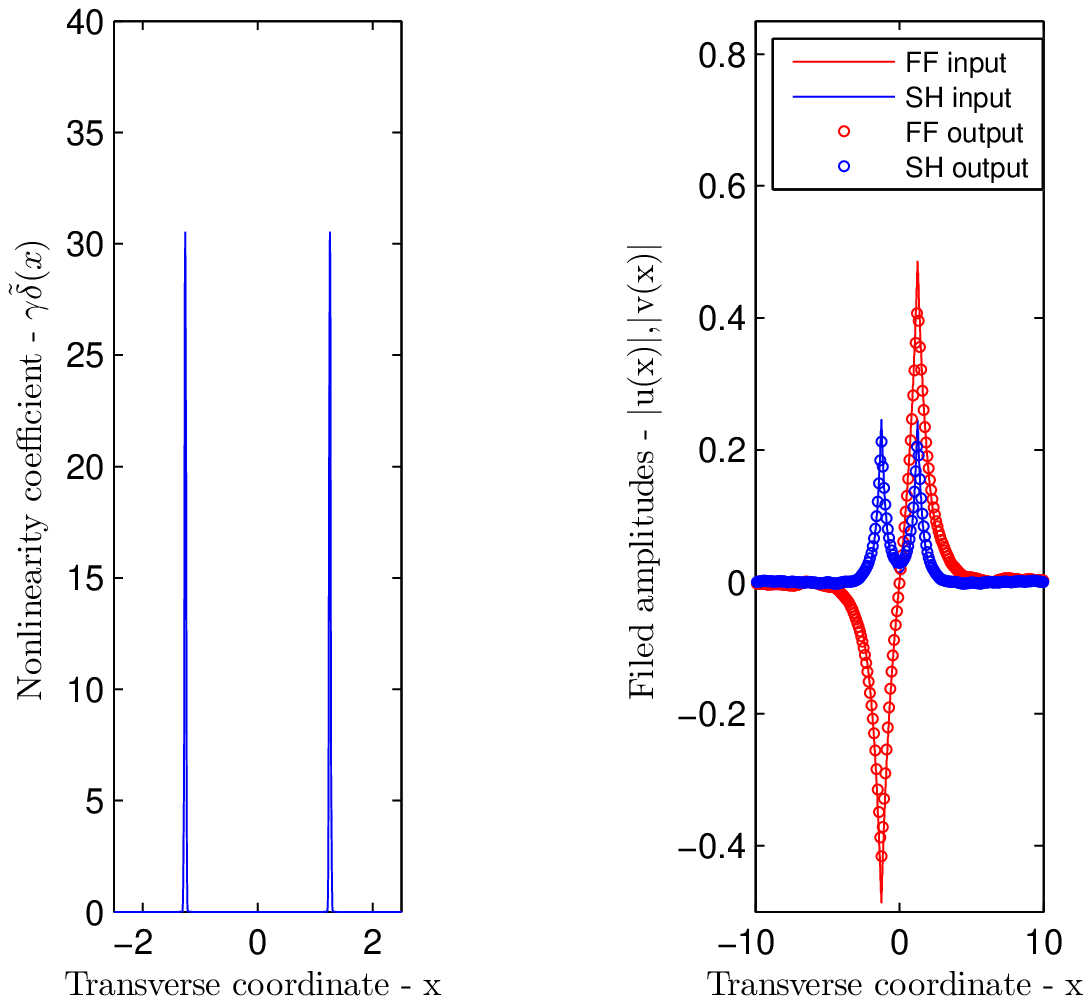}
\caption{(Color online) The same as in Figs. \protect\ref{fig1} and \protect
\ref{fig2}, but for a stable antisymmetric soliton, with $L=200$ $\mathrm{%
\protect\mu }$m, $a=1.4~\mathrm{\protect\mu }$m, and $z=100$ cm.}
\label{fig3}
\end{figure}

Typical examples of stable symmetric, asymmetric and antisymmetric
solitons are displayed in Figs. \ref{fig1}, \ref{fig2}, and \ref{fig3},
respectively. Additional simulations, with strong perturbations added
to the input fields (not shown here), demonstrate that, as it might
be expected, the symmetric solitons are stable before the SSB point
and unstable past it. The simulations also demonstrate that the antisymmetric
solitons are always stable. The stability of asymmetric solitons is
discussed below. %

\begin{figure}[htbp]
 \centering
\includegraphics[width=0.5\textwidth]{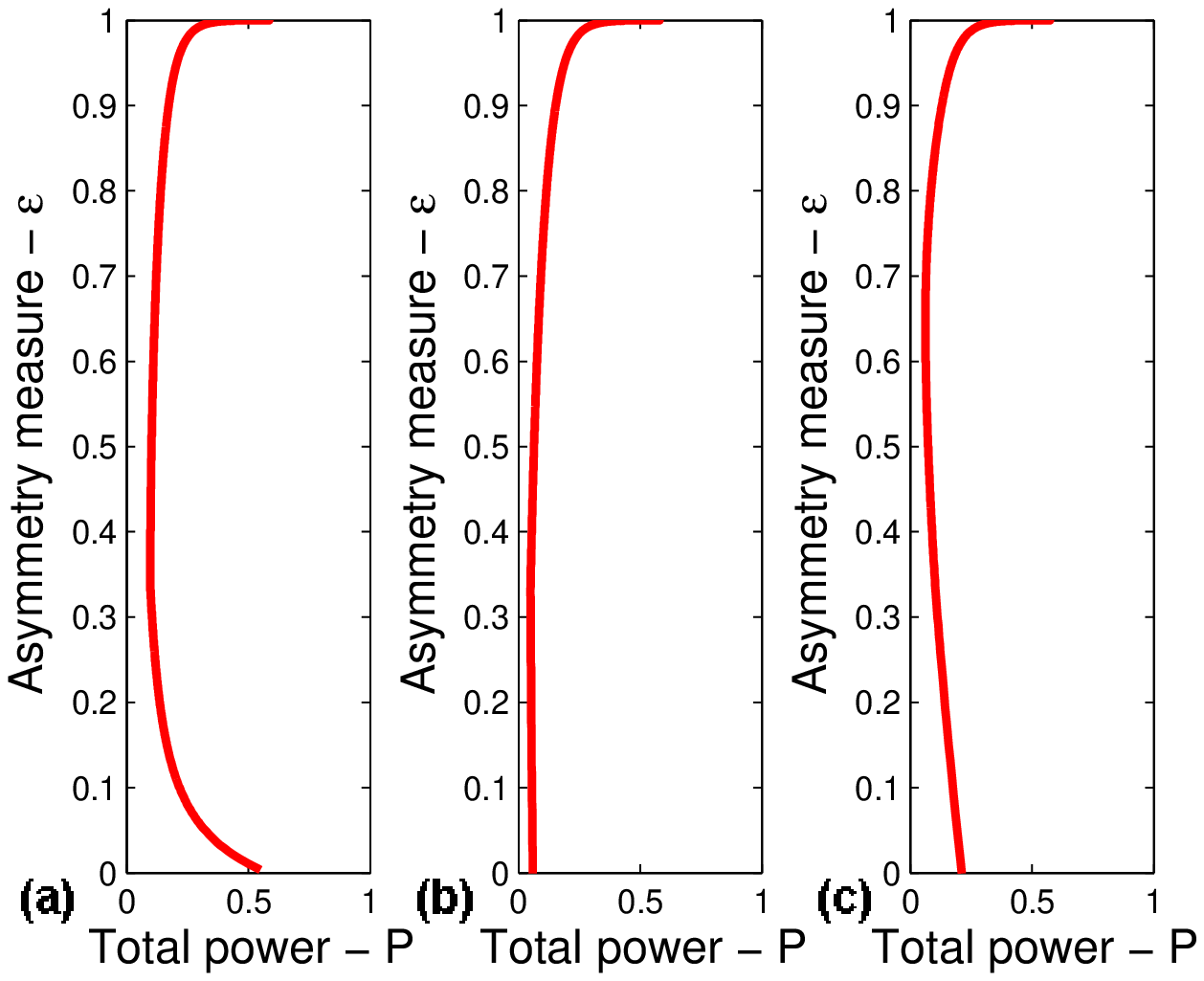}
\caption{(Color online) The asymmetry parameter of the solitons versus the
total power {[}defined as per Eq. (\protect\ref{epsilon}){]}, for positive,
zero, and negative mismatch; $Q=+1$ (a), $Q=0$ (b), and $Q=-1$ (c).}
\label{fig4}
\end{figure}

Making use of Eqs. (\ref{AABB})-(\ref{epsilon}) to find the asymmetry
$\epsilon$, and calculating the total power of the solutions as per
Eq. (\ref{P}), the bifurcation diagrams were drawn in the plane of
$\left(P,\epsilon\right)$, at different values of mismatch $Q$.
Typical examples of the diagrams, displayed in Fig. \ref{fig4}, clearly
demonstrate the \textit{subcritical} character of the SSB bifurcation,
similar to what was found in model (\ref{2delta}) with the cubic
nonlinearity, which corresponds to the cascading limit of the present
system. However, unlike that model, the present one, even with the
ideal $\delta$-functions, is not degenerate, i.e., the branches of
the asymmetric solutions go forward after reaching the turning point.

\begin{figure}[tbp]
\includegraphics[width=0.5\textwidth]{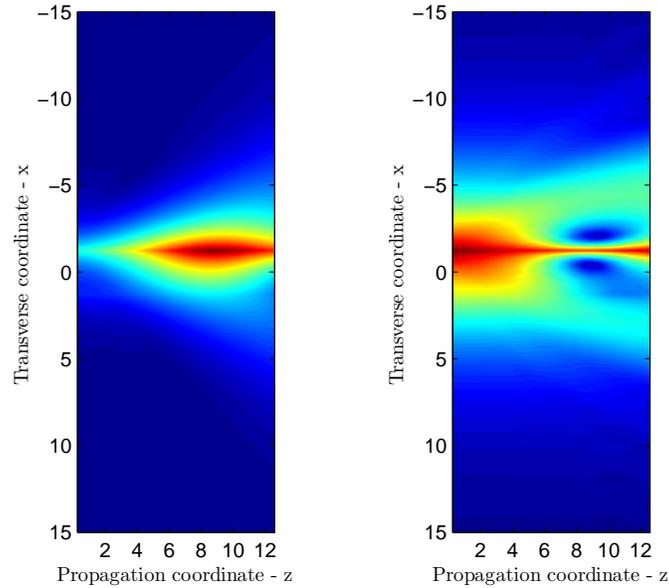}
\caption{(Color online) The evolution of an unstable asymmetric soliton
belonging to the solution branch with $d\protect\epsilon /dP<0$ is displayed
by means of contour plots of local powers of the FF and SH components in the
$\left( z,x\right) $ plane. The corresponding physical parameters are: $%
a=1.4~\mathrm{\protect\mu }$m, $L=200$ $\mathrm{\protect\mu }$m, and the
total propagation distance is $z=100$ cm. The rescaled mismatch and
wavenumber are $Q=-1$ and $k=0.26$.}
\label{fig5}
\end{figure}

In accordance with general properties of the symmetry-breaking bifurcations
\cite{Iooss}, one should expect that branches of the asymmetric solitons
corresponding to $d\epsilon/dP>0$ and $d\epsilon/dP<0$ should be
stable and unstable, respectively. This expectation was confirmed
by direct simulations. In particular, all the asymmetric solitons
belonging to the positive-slope branch of the $\epsilon(P)\ $dependence
are stable (Fig. \ref{fig2} shows an example of such a stable soliton),
while a typical example of the instability of the branches with the
negative slope is displayed in Fig. \ref{fig5}. %

\section{The model with the double nonlinear layer}

A narrow double layer is formed by two adjacent monolayers with opposite
signs of the $\chi^{(2)}$ coefficients. Accordingly, Eqs. (\ref{u})
and (\ref{v}) are replaced by\begin{eqnarray}
iu_{z}+(1/2)u_{xx}+\gamma\delta^{\prime}(x)u^{\ast}v & = & 0,\label{u'}\\
2iv_{z}+(1/2)v_{xx}-Qv+(\gamma/2)\delta^{\prime}(x)u^{2} & = & 0.\label{v'}\end{eqnarray}
An exact stationary solution to Eqs. (\ref{u'}) and (\ref{v'}) can
be found in the form of antisymmetric discontinuous solitons, cf.
expressions (\ref{exact}) for the peakons:\begin{eqnarray}
u\left(z,x\right) & = & \pm\sqrt{2}\gamma^{-1}e^{ikz}\mathrm{sgn}(x)\exp\left(-\sqrt{2k}|x|\right),\nonumber\\
v\left(z,x\right) & = & -\gamma^{-1}e^{2ikz}\mathrm{sgn}(x)\exp\left(-\sqrt{2\left(4k+Q\right)}|x|\right).\label{solution'}\end{eqnarray}
Note that, unlike the peakons, the amplitudes of these solutions do
not depend on $k$.

\begin{figure}[tbp]
\includegraphics[width=0.5\textwidth]{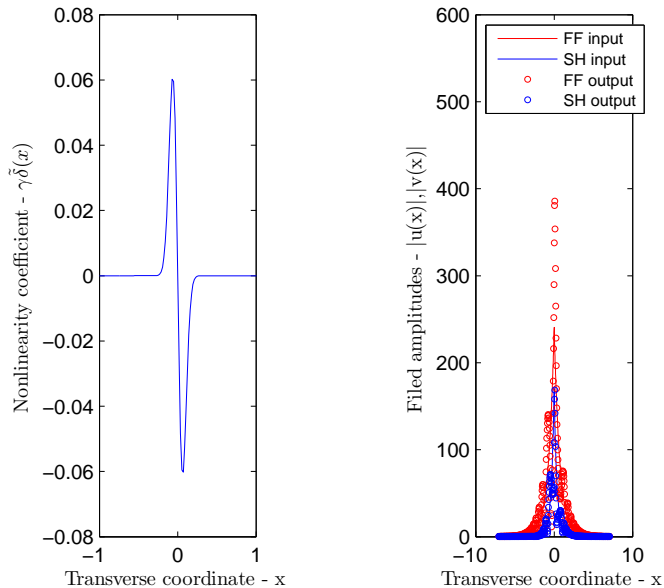}
\caption{(Color online) The same as in Figs. \protect\ref{fig1}-\protect\ref%
{fig3}, but for an unstable antisymmetric nearly discontinuous soliton
pinned to the double nonlinear layer of width $a=7.07~\mathrm{\protect\mu }$%
m, shown in the left panel. In this case, the mismatch is $Q=0$, the total
propagation distance is $1$ cm, and the initial scaled wavenumber is $k=1$.}
\label{fig6}
\end{figure}

According to Eq. (\ref{P}), the total power of the discontinuous
soliton is\begin{equation}
P_{\mathrm{discont}}=\sqrt{2}\gamma^{-1}\left[k^{-1/2}+2\left(4k+Q\right)^{-1/2}\right].\label{T'}\end{equation}
Expression (\ref{T'}) does not give rise to any\emph{\ }existence
threshold, unlike the peakon solutions {[}cf. Eq. (\ref{min}){]},
because $P_{\mathrm{discont}}(k\rightarrow\infty)=0$. Obviously,
Eq. (\ref{T'}) leads to $dP/dk<0$, hence the VK criterion predicts
that the family of the discontinuous solitons is completely \emph{unstable}.
This prediction was confirmed by simulations of the evolution of these
solitons, see Fig. \ref{fig6}, where strong instability dominates
the propagation of the soliton even over a relatively short propagation
distance, $z=1$ cm. %

\section{Nonlinear layers embedded into a nonlinear host medium}

\subsection{A single layer}

If the host medium is itself nonlinear, Eqs. (\ref{u}) and (\ref{v})
are replaced by\begin{eqnarray}
iu_{z}+(1/2)u_{xx}+\left[\Gamma+\gamma\delta(x)\right]u^{\ast}v & = & 0,\label{unonlin}\\
2iv_{z}+(1/2)v_{xx}-Qv+(1/2)\left[\Gamma+\gamma\delta(x)\right]u^{2} & = & 0,\label{vnonlin}\end{eqnarray}
 where $\Gamma$ and $\gamma$ account for the bulk and localized
$\chi^{(2)}$ nonlinearities, respectively, which may have the same
or opposite signs, the latter situation corresponding to the \emph{competing}
bulk and localized nonlinearities. Two particular exact solutions
can be found in this model, following the pattern of the well-known
Karamzin-Sukhorukov (KS) solutions for $\chi^{(2)}$ solitons in the
uniform medium \cite{Moscow}.

\begin{figure}[tbp]
\includegraphics[width=0.5\textwidth]{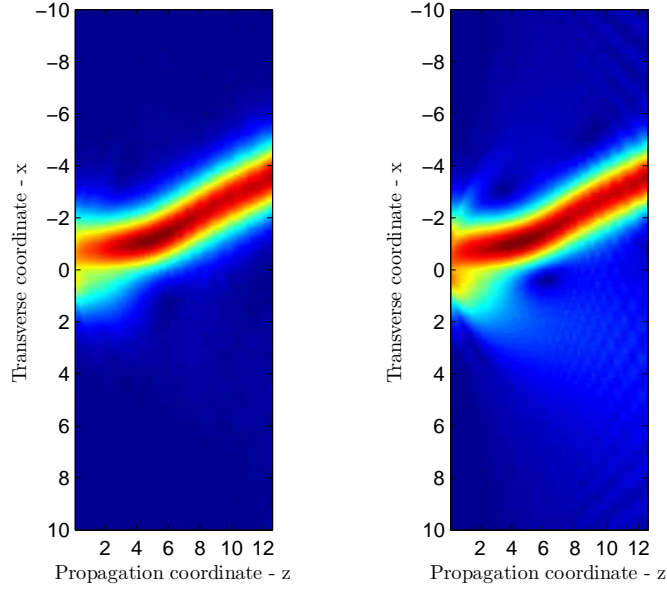}
\caption{(Color online) The same as in Fig. \protect\ref{fig5}, but for the
evolution of the unstable double-humped soliton, given by Eqs. (\protect\ref%
{xicosh})-(\protect\ref{sinhcosh}) with initial scaled wavenumber $k=1$ and $%
Q=-3$, pinned to the nonlinear layer of width $7.07~\mathrm{\protect\mu }$m.%
\textbf{\ }The corresponding total propagation distance is $z=100$ cm.}
\label{fig7}
\end{figure}

The first solution is a straightforward extension of the KS soliton,
based on the following ansatz:\begin{equation}
u=e^{ikz}A~\mathrm{sech}^{2}\left(W\left(|x|+\xi\right)\right),v=e^{2ikz}B\mathrm{sech}^{2}\left(W\left(|x|+\xi\right)\right).\label{xicosh}\end{equation}
Substituting this ansatz into Eqs. (\ref{unonlin}), (\ref{vnonlin})
and taking into regard jump conditions (\ref{ju}), (\ref{jv}), it
is easy to find parameters of the exact solution:\begin{eqnarray}
k=-Q/3,W=\sqrt{-Q/6},A=\pm\sqrt{2}B,~B=-\left(Q/2\Gamma\right),\label{aBA}\\
\sinh\left(2W\xi\right)=\sqrt{-3Q/2}\left(\gamma/\Gamma\right),\label{sinhcosh}\end{eqnarray}
With identical signs of $\gamma$ and $\Gamma$, Eq. (\ref{sinhcosh})
yields $\xi>0$, i.e., a single-hump profile of the pinned soliton
(\ref{xicosh}). For opposite signs of $\gamma$ and $\Gamma$, Eq.
(\ref{sinhcosh}) produces $\xi<0$, hence the corresponding pinned
profile (\ref{xicosh}) features a local minimum at $x=0$, and two
maxima at $x=\pm|\xi|$. The soliton is expected to be stable in the
former case, and unstable in the latter one, when it is pinned by
the repelling defect in an unstable position. Direct simulations of
the evolution of the solitons slightly shifted from the equilibrium
positions confirm these expectations. In particular, the instability
of the double-humped soliton is illustrated by Fig. \ref{fig7}. %

Another type of the pinned soliton can be found in the following form:\begin{equation}
u=e^{ikz}A^{\prime}\left[\sinh\left(W\left(|x|+\xi^{\prime}\right)\right)\right]^{-2},v=e^{2ikz}B^{\prime}\left[\sinh\left(W\left(|x|+\xi^{\prime}\right)\right)\right]^{-2}.\label{xisinh}\end{equation}
The substitution of ansatz (\ref{xisinh}) into Eqs. (\ref{unonlin}),
(\ref{vnonlin}) and (\ref{ju}), (\ref{jv}) produces the following
results, cf. Eqs. (\ref{aBA}) and (\ref{sinhcosh}):\begin{eqnarray}
k=-Q/3,W=\sqrt{-Q/6},A^{\prime}=\pm\sqrt{2}B^{\prime},~B^{\prime}=+Q/\left(2\Gamma\right),\label{aBAsinh}\\
\sinh\left(2W\xi^{\prime}\right)=-\sqrt{-3Q/2}\left(\gamma/\Gamma\right).\label{sinhsinh}\end{eqnarray}
The solution (\ref{xicosh}) is nonsingular provided that Eq. (\ref{sinhsinh})
yields $\xi^{\prime}>0$, which is the case for the \emph{opposite
signs} of $\Gamma$ and $\gamma$.

To better understand the meaning of these exact solutions, it is instructive
to consider the NLS equation which corresponds to the cascading limit
of Eqs. (\ref{unonlin}) and (\ref{vnonlin}):\begin{equation}
iu_{z}+(1/2)u_{xx}+\left[\tilde{\Gamma}+\tilde{\gamma}\delta(x)\right]|u|^{2}u=0,\label{NLS}\end{equation}
$\tilde{\Gamma}\equiv\Gamma^{2}/\left(2Q\right),~\tilde{\gamma}\equiv\gamma\Gamma/Q+\gamma^{2}/\left(2Q\right)\int_{-\infty}^{+\infty}\left[\tilde{\delta}(x)\right]^{2}dx$
{[}recall $\tilde{\delta}(x)$ is the regularized $\delta$-function
(\ref{a}), cf. Eq. (\ref{limit}){]}. In the case of $\tilde{\Gamma}>0$,
i.e., $Q>0$, the exact solution to Eq. (\ref{NLS}), which is the
counterpart of solution (\ref{xicosh}), is\begin{equation}
u=\sqrt{2k/\tilde{\Gamma}}e^{ikz}\mathrm{sech}\left(\sqrt{2k}\left(|x|+\xi\right)\right),\sinh\left(2\sqrt{2k}\xi\right)=2\left(\tilde{\gamma}/\tilde{\Gamma}\right)\sqrt{2k},\label{cosh}\end{equation}
where $k>0$ is the respective wave number. Solution (\ref{cosh})
is valid for both positive and negative $\tilde{\gamma}$, i.e., respectively,
the attractive and repulsive nonlinear defect in Eq. (\ref{NLS}).
The power of solution (\ref{cosh}) is\begin{equation}
P=\int_{-\infty}^{+\infty}\left\vert u(x)\right\vert ^{2}dx=|\tilde{\gamma}|^{-1}\left[\left(2\left\vert \tilde{\gamma}\right\vert /\tilde{\Gamma}\right)\sqrt{2k}-\mathrm{sgn}\left(\Gamma\right)\sqrt{8\left(\tilde{\gamma}/\tilde{\Gamma}\right)^{2}k+1}+1\right].\label{Ncosh}\end{equation}
It immediately follows from this expression that, for either sign
of $\tilde{\gamma}$, this soliton family satisfies the VK criterion,
$dP/dk>0$, which suggests that solution (\ref{xicosh}) (with identical
signs of $\gamma$ and $\Gamma$) is stable in the general case too,
when the cascading limit does not apply. As mentioned above, this
expectation was corroborated by direct simulations (not shown here).

\begin{figure}[tbp]
\includegraphics[width=0.5\textwidth]{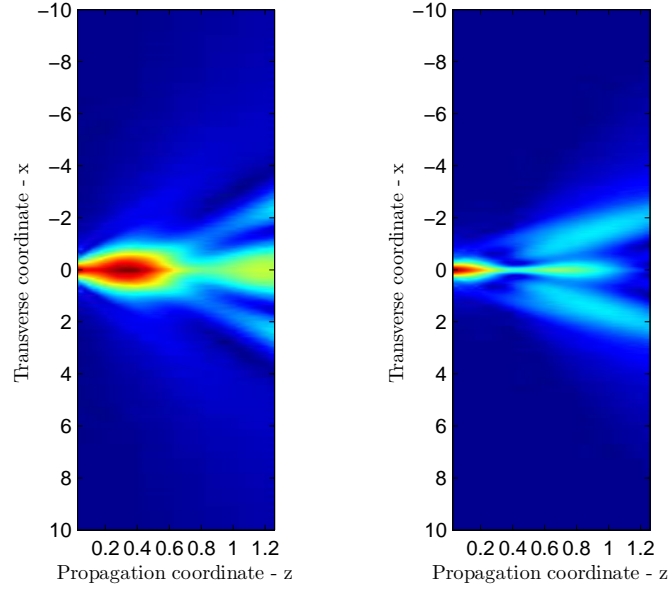}
\caption{(Color online) The same as in Fig. \protect\ref{fig7}, but for the
evolution of the unstable soliton, given by Eqs. (\protect\ref{xisinh})-(%
\protect\ref{sinhsinh}) with initial wavenumber $k=1$ and $Q=-3$, pinned to
the nonlinear layer of width $7.07~\mathrm{\protect\mu }$m.\textbf{\ }The
corresponding total propagation distance is $z=10$ cm.}
\label{fig8}
\end{figure}

The cascading-limit counterpart of solution (\ref{xisinh}) corresponds
to $\tilde{\Gamma}<0$, $\tilde{\gamma}>0$. The respective exact
solution to Eq. (\ref{NLS}) and its power are\begin{equation}
u=\sqrt{2k/|\tilde{\Gamma}|}e^{ikz}\left[\sinh\left(\sqrt{2k}\left(|x|+\xi\right)\right)\right]^{-1},\sinh\left(2\sqrt{2k}\xi\right)=2\left(\tilde{\gamma}/|\tilde{\Gamma}|\right)\sqrt{2k},\label{sinh}\end{equation}
\begin{equation}
P=\tilde{\gamma}^{-1}\left[\sqrt{8\left(\tilde{\gamma}/\tilde{\Gamma}\right)^{2}k+1}-\left(2\tilde{\gamma}/\tilde{\Gamma}\right)\sqrt{2k}+1\right].\label{Nsinh}\end{equation}
Expression (\ref{Nsinh}) \emph{does not} satisfy the VK criterion,
as it yields $dP/dk<0$, suggesting an instability of solution (\ref{xisinh})
in the absence of the cascading limit. Indeed, direct simulations
of Eqs. (\ref{unonlin}) and (\ref{vnonlin}) confirm that this solution
is unstable, as shown in Fig. \ref{fig8}. %

\subsection{A linear layer embedded into the self-defocusing nonlinear medium
(the cascading limit)}

The existence of solution (\ref{sinh}) to the asymptotic NLS equation
(\ref{NLS}) suggests to consider a similar solution for a \emph{linear}
attractive layer (i.e., a usual waveguiding channel) embedded into
the medium with the uniform self-defocusing cubic nonlinearity (in
previous works, solitons pinned by the attractive defect were considered
in the NLS equation with the self-focusing nonlinearity \cite{Cao,Weinstein}).
In terms of the $\chi^{(2)}$ system, this solution corresponds to
the narrow linear channel in the limit of the large \emph{negative}
mismatch. The respective version of the NLS equation is\begin{equation}
iu_{z}+(1/2)u_{xx}+\Gamma|u|^{2}u+\gamma_{0}\delta(x)u=0,\label{linear}\end{equation}
with $\Gamma<0$ and $\gamma_{0}>0$. In fact, we can fix $\Gamma\equiv-1$
in this case; then, an exact solution to Eq. (\ref{linear}) for a
mode pinned by the attractive layer, is {[}cf. Eqs. (\ref{sinh}){]}\begin{equation}
u=\sqrt{2k}e^{ikz}\left[\sinh\left(\sqrt{2k}\left(|x|+\xi\right)\right)\right]^{-1},\tanh\left(\sqrt{2k}\xi\right)=\sqrt{2k}/\gamma_{0},\label{sinhlin}\end{equation}
with power $P=\gamma_{0}-\sqrt{2k}$, which features $dP/dk<0$, formally
contradicting the VK criterion. However, this criterion is not relevant
for models with the self-defocusing nonlinearity. Actually, an argument
in favor of the stability of solution (\ref{sinhlin}) is the fact
that its energy is negative,\begin{eqnarray}
E & \equiv & \frac{1}{2}\int_{-\infty}^{+\infty}\left[\left(u_{x}\right)^{2}-u^{4}\right]dx-\gamma_{0}u^{2}\left(x=0\right)\nonumber \\
 & = & -\left(\sqrt{2}/3\right)k^{3/2}\left(\gamma_{0}/\sqrt{2k}-1\right)\left(2\gamma_{0}/\sqrt{2k}+1\right)^{2}\end{eqnarray}
(recall $\Gamma=-1$ was fixed), hence the solution has a good chance
to represent the \textit{ground state} of the system. Direct simulations
of Eq. (\ref{linear}) corroborate the stability of this pinned mode
(not shown here).

\subsection{Two embedded layers}

The extension of the model for the pair of symmetric layers is described
by the following version of Eqs. (\ref{unonlin}) and (\ref{vnonlin})
{[}cf. Eqs. (\ref{2u}) and (\ref{2v}) in the case of the linear
host medium{]}:\begin{eqnarray}
iu_{z}+\frac{1}{2}u_{xx}+\left\{ \Gamma+\gamma\left[\delta\left(x-\frac{L}{2}\right)+\delta\left(x+\frac{L}{2}\right)\right]\right\} u^{\ast}v & = & 0,\label{u2nonlin}\\
2iv_{z}+\frac{1}{2}v_{xx}-Qv+\frac{1}{2}\left\{ \Gamma+\gamma\left[\delta\left(x-\frac{L}{2}\right)+\delta\left(x+\frac{L}{2}\right)\right]\right\} u^{2} & = & 0.\label{v2nonlin}\end{eqnarray}
A particular exact solution to Eqs. (\ref{u2nonlin}) and (\ref{v2nonlin})
can be found, in the form of a symmetric \textit{three-hump} structure
{[}with a maximum at $x=0$, on the contrary to the double-humped
solution (\ref{genu}), (\ref{genv}), which has a minimum at $x=0${]},
for the competing nonlinearities, $\Gamma>0$, $\gamma<0$:\begin{equation}
u(x,z)=Ae^{-ikz}\left\{ \begin{array}{c}
\left[\cosh\left(\sqrt{-Q/6}\left(|x|-L\right)\right)\right]^{-2},~~\mathrm{at}~~|x|>L/2,\\
\left[\cosh\left(\sqrt{-Q/6}x\right)\right]^{-2},~~\mathrm{at}~~|x|<L/2,\end{array}\right.\label{u3hump}\end{equation}
\begin{equation}
v(x,z)=Be^{-2ikz}\left\{ \begin{array}{c}
\left[\cosh\left(\sqrt{-Q/6}\left(|x|-L\right)\right)\right]^{-2},~~\mathrm{at}~~|x|>L/2,\\
\left[\cosh\left(\sqrt{-Q/6}x\right)\right]^{-2},~~\mathrm{at}~~|x|<L/2,\end{array}\right.\label{v3hump}\end{equation}
where $k$, $A$ and $B$ are given by the same expressions (\ref{aBA})
as in the case of solution (\ref{xicosh}).

\begin{figure}[tbp]
\includegraphics[width=0.5\textwidth]{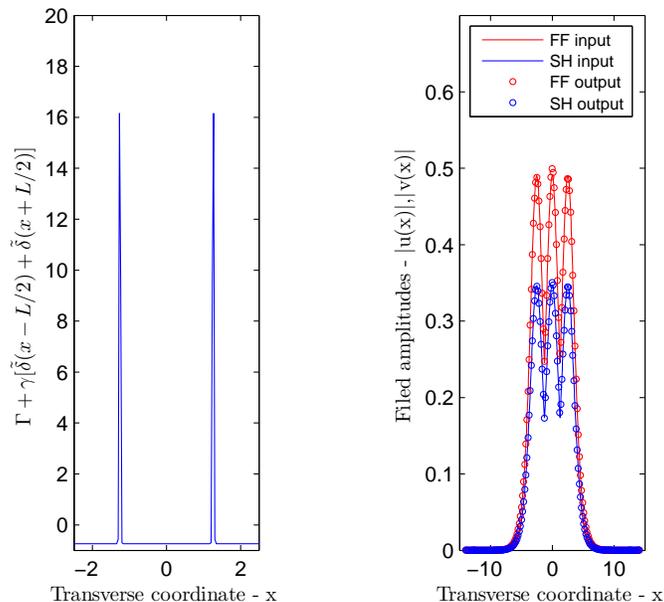}
\caption{(Color online) Left: the profile of the nonlinearity coefficient,
for separation $L=200$ $\mathrm{\protect\mu }$m between two symmetric layers
of width $a=1.77~\mathrm{\protect\mu }$m each. Right: a stable three-humped
solution given by Eqs. (\protect\ref{u3hump}), (\protect\ref{v3hump}), for $%
Q=-3$, $k=1$, and $z=100$ cm.}
\label{fig9}
\end{figure}

The highly degenerate nature of this solution is demonstrated by the
fact that it satisfies the jump conditions (\ref{ju}), (\ref{jv})
at points $x=\pm L/2$ at a single value of the strength of the localized
nonlinearity, $\gamma=\left(2\Gamma/\sqrt{-6Q}\right)\sinh\left(\sqrt{-Q/6}L\right)$
(recall that both $Q$ and $\gamma$ are negative, while $\Gamma$
is positive, in the present case). As for the stability of the three-hump
mode, it may be expected that, with a maximum of the local power set
between two \emph{repulsive} nonlinear layers,\ it is definitely
unstable at small distance $L$ between the layers (when they tend
to merge into a single repulsive element, cf. the instability in Fig.
\ref{fig7}), but it may become stable at larger $L$, when the repulsion
from the two separated layers traps the power maximum between them.
These expectations have been corroborated by direct simulations of
Eqs. (\ref{u2nonlin}) and (\ref{v2nonlin}). In particular, an example
of the stable mode, for large separation $L=200~\mathrm{\mu}$m, is
displayed in Fig. \ref{fig9} (this solution is stable against large
perturbations, which is not shown here in detail). %

\section{Conclusion}

We have produced several exact solutions for spatial solitons supported
by the $\chi^{(2)}$ layers embedded into a linear or nonlinear planar
waveguide. The most fundamental solution describes the full set of
families of the symmetric, asymmetric, and antisymmetric double-humped
modes supported by the symmetric pair of the nonlinear layers inserted
into the linear medium. The exact solutions describe the subcritical
symmetry-breaking bifurcation in this system. In addition, particular
exact solutions of several types were found for nonlinear stripes
running through the nonlinear medium, including the case of the competition
between the uniform and localized $\chi^{(2)}$ nonlinearities. The
stability of the pinned solitons was tested by means of direct numerical
simulations. In the case of the pair of nonlinear stripes embedded
into the linear waveguide, the character of the (in)stability completely
agrees with general principles of the bifurcation theory. For the
layers embedded into the nonlinear host medium, the results for the
stability were explained too, with the help of the consideration of
the cascading limit.

The theoretical results reported in this paper call for an experimental
realization. As an example, for a small phase mismatch $\triangle k=20$
m$^{-1}$ 
and a typical nonlinear coefficient of $13~$ pm/V, the input intensities
required to observe the soliton in the case of the single embedded
layer are about $10^{9}$~
W/cm$^{2}$. Such intensities are feasible, as demonstrated in Ref.
\cite{Saltiel}, provided that the necessary nonlinear pattern can
be fabricated.

\bigskip{}
 {} N.V.B. is an Eshkol Scholar from the Israeli ministry of science,
culture and sport.

\end{document}